\documentclass[12pt]{article}
\usepackage{amsmath,amssymb}
\usepackage{graphicx,color}
\numberwithin{equation}{section}
\usepackage{cite}
\usepackage{bm}
\usepackage{dcolumn}
\newcommand{\bea}{\begin{eqnarray}}
\newcommand{\eea}{\end{eqnarray}}
\newcommand{\nn}{\nonumber}
\newcommand{\na}{\nabla}
\begin{document}

\vspace{12mm}

\begin{center}
{{{\Large {\bf A
New Extension of
%Anisotropic Weyl-invariant
Ho\v{r}ava-Lifshitz
Gravity and Curing Pathologies of the Scalar Graviton
%-invariant
%Weyl-Extended
%Horava-Lifshitz
 }}}}\\[10mm]

{Taeyoon Moon$^{a}$\footnote{e-mail address: dpproject@skku.edu},
Phillial Oh$^{b}$\footnote{e-mail address: ploh@skku.edu}, and Mu-In
Park$^{c}$\footnote{e-mail address: muinpark@gmail.com}
}\\[8mm]

{\em {${}^{a}$ Center for Quantum Space-time,\\
Sogang University, Seoul, 121-742, Korea\\[7pt]
${}^{b}$ Department of Physics and Institute of Basic Science,\\
Sungkyunkwan University, Suwon 440-746 Korea}\\[7pt]
${}^{c}$ The Institute of Basic Sciences,\\
Kunsan National University, Kunsan, 573-701, Korea\\[0pt]
}

\end{center}
\vspace{2mm}

\begin{abstract}
We consider an extension of the Ho\v{r}ava-Lifshitz gravity with
extra
conformal symmetry by introducing a scalar field with %the
higher order curvature terms. Relaxing the exact local Weyl
symmetry, we construct an action with three free parameters which
breaks local anisotropic Weyl symmetry but still preserves residual
global Weyl symmetry. At low energies, it reduces to a
Lorentz-violating scalar-tensor gravity. With a constant scalar
field background and particular choices of the parameters, it
reduces to the Ho\v{r}ava-Lifshitz (HL) gravity, but any
perturbation from these particular configurations produces some
non-trivial extensions of HL gravity. The perturbation analysis of
the new extended HL gravity in the Minkowski background shows that
the pathological behaviors of scalar graviton, i.e., ghost or
instability problem, and strong coupling problem do not emerge up to
cubic order as well as quadratic order.

\end{abstract}
\vspace{5mm}


%{\footnotesize ~~~~Keywords: Classical Theories of Gravity, Spacetime Singularities, Black Holes in String Theory}

\vspace{1.5cm}

\hspace{11.5cm}{Typeset Using \LaTeX}
\newpage
\renewcommand{\thefootnote}{\arabic{footnote}}
\setcounter{footnote}{0}
\section{Introduction}

~~~~~The renormalizability of Einstein gravity has been a long
standing problem in quantum gravity. The power counting
renormalizability of curvature-squared gravity was obtained by
adding the most general, covariant, higher-derivative action
containing only dimensionless couplings to the Einstein-Hilbert
action with a cosmological constant term \cite{utiyama}. Rigorous
renormalizability was established by using BRS invariance
\cite{stelle}. Also, conformal gravity with Weyl tensor was  first
considered by Bach \cite{bach} in 1921. Subsequently, it has been
shown that this theory is renormalizable and asymptotic free
\cite{fradkin, riegert}. However, these curvature-squared gravities
have the pathologies of ghost and unitarity problem.
For example, conformal gravity %propagates
have two ghost modes consisting of a massless vector and a massless
runaway tensor. The four extra ghost degrees of freedom are due to
the higher-derivative nature \cite{pais}.

Recently, Ho\v{r}ava \cite{Horava} proposed a renormalizable theory
of quantum gravity which is known as Ho\v{r}ava-Lifshitz (HL)
gravity. The basic idea is to abandon the Lorentz invariance and
equal-footing treatments of space and time in UV. This theory is an
Einstein gravity with a Lorentz violating parameter $\lambda$ which
reduces to the usual Einstein gravity with $\lambda=1$ at low (IR)
energies. It is power-counting renormalizable without ghosts for the
usual transverse traceless graviton mode and contains $6$th-order
spatial derivatives. It is based on the detailed balance and the
projectability which restrict the lapse function to be a function of
time only, $N=N(t)$.

The original HL gravity with the projectability is known to have a
few serious problems. These are the existence of the extra degree of
freedom of graviton ({\em scalar graviton}) and the strong coupling
problem at the Einstein gravity limit, i.e., $\lambda \rightarrow
1$, in IR \cite{chr,blas2}. As was shown in \cite{koyama}, the
scalar graviton can be a ghost or leads to instability and moreover
the couplings of cubic order terms blow up at the Einstein gravity
limit (strong coupling) in the Minkowski background that makes the
metric perturbation break down for the scalar graviton. The strong
coupling might not be a problem but only means the necessity of
non-perturbative analysis for the Einstein gravity limit. Actually,
it was argued in \cite{izmi,mukohyama} that this strong coupling
problem can be eliminated through a non-perturbative effect, like
the Vainshtein mechanism in massive gravity \cite{vain}.

 There are two main alternative models of extended HL gravity. The
first one was suggested by Sotiriou, Visser, and Weinfurtner (SVW)
\cite{sot} and the other is the so called healthy extension of HL
gravity by Blas,  Pujolas, and Sibiryakov (BPS) \cite{Blas}. SVW
 were motivated by the fact that HL gravity had a non-zero cosmological
 constant of the wrong sign to be
incompatible with observation. To overcome the problem, firstly they
have constructed the gravity model by abandoning the detailed
balance condition and restoring parity invariance but with
projectability. As pointed out in \cite{sot,sotiriou}, SVW approach
has still the pathology of the scalar graviton. (For related issues
in de Sitter background, see \cite{wang,Wang}.)

On the other hand the motivation of BPS was to improve the IR
behavior without detailed balance condition and projectability. To
do so they first introduced a new $3$-vector $a_i=\partial_i N/N$
and its higher derivative terms into the Lagrangian. Here, $N$
became a dynamical scalar field and at low energies, it reduced to a
Lorentz-violating scalar-tensor gravity theory. This new model
endowed the scalar graviton with a regular action. Consequently the
pathology of scalar graviton, such as ghost or instability problem
can be cured in BPS extension but it is known that this extension
also could have strong coupling problem at the Einstein limit
($\lambda \rightarrow 1$) in IR when one considers cubic order
action \cite{sotiriou2,Kimp}. (See also \cite{Soti:1103} for
low-dimensional analogous.) However, it is also possible to avoid
the strong coupling if higher-derivative terms in the action become
important below the strong coupling energy scale \cite{blas3,blas4},
by assigning hierarchy between the Planck scale and a new low energy
scale.

Recently, another extension of HL gravity was performed with a
conformally invariant manner and the local anisotropic Weyl gravity
was constructed \cite{jcap}. It extends the original anisotropic
Weyl invariance of HL gravity at UV to that of all energy scales
using an extra scalar field which compensates the local scale
transformation. The action is invariant under the local anisotropic
transformations of the space and time metric components with an
arbitrary value of the critical exponent $z$. It turns out that this
theory coincides with the low-energy limit of the non-projectable HL
gravity and it permits the extra scalar graviton mode which
%naturally
inherits the pathologies of the HL gravity.

%In this paper, we consider conformal extension of Horava-Lifshitz
%gravity and construct anisotropic Weyl gravity including higher
%derivative terms. Then, we discuss the scalar graviton and strong
%coupling problems of HL gravity in this framework.
In this paper, we relax the exact local conformal invariance and
consider anisotropic $z=3$ Weyl gravity including the higher order
derivative terms. It includes three
%five({\bf ?})
parameters which represents the breaking of the local Weyl
invariance, but still preserves the global conformal invariance.
When all of these parameters become $1$, the theory has local
invariance. With fixing the scalar field to a constant value and
%suitable
particular choices of the parameters, the theory reduces to the HL
gravity, but any perturbation from these particular configurations
produces some non-trivial extensions of HL gravity. We study the
behaviors of the scalar graviton in the perturbation analysis of
this new extended HL gravity and show that, in the Minkowski
background, the pathological behaviors of scalar graviton, i.e.,
ghost or instability problem, and strong coupling problem do not
emerge up to cubic order as well as quadratic order.

\section{Anisotropic Weyl-invariant action with higher derivatives:
New extended HL gravity }

~~~~~In order to construct anisotropic
Weyl-invariant action with higher derivative %order
terms, let us first consider $z=3$ anisotropic Weyl-invariant
gravity \cite{jcap}
\begin{eqnarray}
S_{aW}=\int \,dt
d^3xN\sqrt{g}~\left\{\frac{2}{\kappa^2}\left(B_{ij}B^{ij} -\lambda
B^2\right)+\varphi^{8}\left(R-8
\frac{\nabla_{i}\nabla^{i}\varphi}{\varphi}\right)- V_{\nu}(\varphi)
%\nu \varphi^{12}
\right\}, \label{aW}
\end{eqnarray}
where $\kappa^2,~\lambda$ are dimensionless constant parameters, $R$
is the $3$-curvature, and $B_{ij}$ is given by
\begin{eqnarray}
B_{ij}=K_{ij}-\frac{2}{N\varphi}g_{ij}(\dot{\varphi}-\na_{i}\varphi
N^{i}),
%\\&\equiv&K_{ij}+\frac{\theta}{2N} g_{ij},
\end{eqnarray}
with
%$\theta=-4(\dot{\varphi}-\nabla_{i}\varphi N^{i})/\varphi$ and
the
extrinsic curvature $K_{ij}= -(\dot{g}_{ij}-\nabla_i N_j-\nabla_jN_i
)/2N$ (the dot $(\dot{~})$ denotes the derivative with respect to
$t$). One can easily check that for
\begin{eqnarray}
V_{\nu}(\varphi)=\nu \varphi^{12}
\end{eqnarray}
 (with a constant coefficient $\nu$) the above action
(\ref{aW}) is invariant under anisotropic Weyl transformation
\begin{eqnarray}
 N \rightarrow
e^{3\omega}N,~~~N_{i}\rightarrow e^{2\omega}N_{i},%\nn\\
~~~g_{ij} \rightarrow e^{2\omega}g_{ij},~~\varphi\rightarrow
e^{-\frac{\omega}{2}}\varphi.\label{trans}
\end{eqnarray}
In these transformations, $\omega$ is a function of space and time,
$\omega=\omega(t, {\bf x})$. Note that assuming $\omega$ as a
function of time only, i.e., $\omega=\omega(t)$, is unnatural from
the point of view of the above local transformations. This implies
that the lapse function $N$ must be a function of space and time
also, and this favors the non-projectable case in our construction.
Later, we will consider breaking of the above local anisotropic Weyl
invariance but keep only the global invariance so that the
projectable  case is still possible. However, even in this case we
will consider only the non-projectable case in order to study
whether the scalar graviton problem in the BPS extension
%gravity
\cite{sotiriou2,Kimp} can be cured in our new construction.
%We find that depending on the values of the symmetry
%breaking parameters, the non-projectability has to be taken.

We also note that, for $\lambda=1$ and
$\varphi=\varphi_0=\mbox{const.}$, the action (\ref{aW}) is reduced
to (Lorentz-invariant) Einstein-Hilbert action with
%the cosmological constant ($\Lambda$) and
the following conditions \cite{Park:0906}:
\begin{eqnarray}
\frac{2}{\kappa^2}=\frac{c^2}{16\pi G_N
},~~~\varphi_0^8=\frac{c^4}{16\pi G_N},~~~V_{\nu}(\varphi_0)=\nu
 \varphi_0^{12}=\frac{2\Lambda c^2}{16\pi G_N},~~~ \nu=\frac{\sqrt{16 \pi G_N} 2
\Lambda}{c^4} ,\label{Elimit}
\end{eqnarray}
where $G_N$ is the gravitational constant, $c$ is the speed of
light, and $\Lambda$ is the cosmological constant. However, we
should point out that,
%from (\ref{Elimit}) Einstein limit with $\lambda=1$ can be
%achieved independent of the cosmological constant term ($\Lambda$).
for $\lambda \neq 1$ or $\varphi^8 \neq {c^4}/{16\pi G_N}$, the
action is not invariant under the full diffeomorpism ({\it Diff})
but invariant under the foliation preserving {\it Diff}:
\begin{eqnarray}
\label{Diff}
\delta x^i &=&-\zeta^i (t, {\bf x}), ~\delta t=-f(t), \nonumber \\
 \delta
g_{ij}&=&\partial_i\zeta^k g_{jk}+\partial_j \zeta^k g_{ik}+\zeta^k
\partial_k g_{ij}+f \dot g_{ij},\nonumber\\
\delta N_i &=& \partial_i \zeta^j N_j+\zeta^j \partial_j
N_i+\dot\zeta^j
g_{ij}+f \dot N_i+\dot f N_i, \nonumber \\
\delta N&=& \zeta^j \partial_j N+f \dot N+\dot f N, \nonumber\\
\delta \varphi &=&\zeta^k
\partial_k \varphi+f \dot \varphi .
\end{eqnarray}
Here it is important to note that this {\it Diff} exists for
arbitrary spacetime-dependent $N,N_i,g_{ij}, \varphi$. This implies
that the equations of motion by varying $N,N_i,g_{ij},\varphi$ are
all the ``local'' equations as in the usual Lorentz invariant
Einstein or scalar-tensor gravity. This is compatible with the local
Weyl invariance (\ref{trans}). So, there are two sources of the IR
Lorentz violation: One comes from the parameter $\lambda \neq 1$ and
another from any fluctuation of $\varphi$ from the background
$\varphi_0^8 ={c^4}/{16\pi G_N}$.

When we focus on the following Weyl invariant object:
\begin{eqnarray}
\bar{R}_{ij}&\equiv&R_{ij}+6\frac{\nabla_i\varphi\nabla_j\varphi}{\varphi^{2}}
-2\frac{\nabla_i\nabla_j\varphi}{\varphi}
-2g_{ij}\frac{\nabla_k\varphi\nabla^k\varphi}{\varphi^{2}}
-2g_{ij}\frac{\nabla_k\nabla^k\varphi}{\varphi} \nonumber \\
&\equiv& R_{ij}+f_{ij}(\nabla\varphi),~~~\bar{R}\equiv
g^{ij}\bar{R}_{ij}
\end{eqnarray}
we can further extend $S_{aW}$ (\ref{aW}) to the power-counting
renormalizable and local Weyl invariant action including the
higher-derivative %order
terms as
\begin{eqnarray}
&&\hspace*{-2em}S_{haW}=\int \,dt
d^3xN\sqrt{g}~\Bigg\{\frac{2}{\kappa^2}\left(B_{ij}B^{ij} -\lambda
B^2\right)-V_{\nu}(\varphi)+\varphi^{8}\bar{R}+
\beta_1\varphi^{4}\bar{R}^2+\beta_2\varphi^4(\bar{R}_{ij})^2
\nn\\
&&\hspace*{3em} +\beta_3 \bar{R}^3+ \beta_4\bar{R}(\bar{R}_{ij})^2
+\beta_5\bar{R}_{ij}\bar{R}^{jk}\bar{R}_{k}^i+\beta_6\bar{\nabla}_{i}\bar{R}_{jk}
\bar{\nabla}^{i}\bar{R}^{jk}+\beta_7(\bar{\nabla}_i
\bar{R})^2\Bigg\},\label{haW}
\end{eqnarray}
where $\beta_{1\sim 7}$ are arbitrary constant parameters and
$\bar{\nabla}_{i}\bar{R}_{jk}=\nabla_{i}\bar{R}_{jk}-\Psi_{ij}^{~~l}\bar{R}_{lk}
-\Psi_{ik}^{~~l}\bar{R}_{jl}$ with
$\Psi_{ij}^{~~l}=-2\varphi^{-1}\left(\nabla^l \varphi
g_{ij}-\nabla_i \varphi\delta_{j}^{l}-\nabla_j
\varphi\delta_{i}^{l}\right)$.
%Especially for $\lambda=1/3$, $B_{ij}B^{ij}-B^2/3$ term can be
%reduced to $K_{ij}K^{ij}-K^2/3$ term. This means that the above
%action becomes anisotropic Weyl invariant Horava-Lifshitz action for
%special choice of the parameters, will be shown below.
 Note that here one can always choose %also that with
a gauge
%choice of
$\varphi=\mbox{const}$ as in (\ref{Elimit}) such that
%, the
%terms of 3rd and 4th line in (\ref{haW}) are reduced to the cubic
%terms in the literature
it reduces to the SVW action \cite{sot}. Then, the physical contents
of the original SVW action,
%gravity,
like as the scalar graviton problem, would be the same in this
extended gravity also. So, there would be no fundamental advantage
of this extension to resolve the scalar graviton problem of the
original SVW approach.
%gravity.

For this reason, we %can
consider %the
a new extended action with three deformation parameters
($\xi_1,\xi_2,\xi_3$) which break, for $\xi_{1\sim3}\neq 1$, the
%ing
local anisotropic Weyl invariance
%symmetry
but still preserve
%s
residual global Weyl invariance as follows:
\begin{eqnarray}
&&\hspace*{-2em}S=\int \,dt
d^3xN\sqrt{g}~\Bigg\{\frac{2}{\kappa^2}\left(\tilde{B}_{ij}\tilde{B}^{ij}
-\lambda \tilde{B}^2\right)-V_{\nu}(\varphi)+\varphi^{8}\tilde{R}+
\beta_1\varphi^{4}\tilde{R}^2+\beta_2\varphi^4(\tilde{R}_{ij})^2
\nn\\
&&\hspace*{3em} +\beta_3 \tilde{R}^3+
\beta_4\tilde{R}(\tilde{R}_{ij})^2
+\beta_5\tilde{R}_{ij}\tilde{R}^{jk}\tilde{R}_{k}^i+\beta_6\bar{\nabla}_{i}\tilde{R}_{jk}
\bar{\nabla}^{i}\tilde{R}^{jk}+\beta_7(\bar{\nabla}_i
\tilde{R})^2\Bigg\},\label{Action}
\end{eqnarray}
where
%$\tilde{B}_{ij}=K_{ij}+\frac{\tilde\theta}{2N}g_{ij}$ with
%$\tilde\theta=-4(\xi_1\dot{\varphi}-\xi_2\nabla_{i}\varphi
%N^i)/\varphi$ and
\begin{eqnarray}
\tilde{B}_{ij}&\equiv&K_{ij}-\frac{2}{N\varphi}g_{ij}(\xi_1\dot{\varphi}-\xi_2\na_{i}\varphi
N^{i}), \\
%\\&\equiv&K_{ij}+\frac{\theta}{2N} g_{ij}, \\
\tilde{R}_{ij}&\equiv& R_{ij}+\xi_3
f_{ij}(\nabla\varphi),~~~\tilde{R}=g^{ij} \tilde{R}_{ij}.
\end{eqnarray}
%In the above action, the choice of
%$\xi_{1\sim3}=1,~V(\varphi)\sim\varphi^{12}$ corresponds to the
%anisotropic Weyl invariance under the transformations (\ref{trans}).
%It is important to note that
Here, the parameter $\xi_1$ is associated with the breaking of local
Weyl invariance along the time slice of the extrinsic curvature
scalar. Whereas $\xi_{2}$ and $\xi_{3}$ are associated with the
non-invariances along the spatial directions of the extrinsic
curvature and $3$-dimensional curvature scalar, respectively.
%This is not the most general action with broken Weyl invariance
 Note that the local Weyl invariance is not completely broken, but
there are some residual local symmetries left, depending on the
parameters; for example, when
%$(\xi_1, \xi_2, \xi_3)=(1, \neq 1, \neq 1)$
$\xi_1=1$, $\xi_{2,3}\neq 1$, the transformation function $\omega$
can be an arbitrary function of time;  for
%$(\xi_1, \xi_2, \xi_3)=(\neq 1, 1, 1)$
$\xi_1\neq 1$, $\xi_{2,3}= 1$, $\omega$ can be  an arbitrary
function of space.

 It is important to note that, due to lack of local Weyl invariance,
one can not choose the ``gauge'' $\varphi=\varphi_0=\mbox{const}$
always to reduce the theory to the HL gravity. This means that there
is the additional, physical, scalar degree of freedom $\varphi$.
But, in the absence of the scalar fluctuation mode around the
background $\varphi_0$, this theory should be reduced to the
Einstein-Hilbert action in IR such that the conditions in
(\ref{Elimit}) is to be satisfied again\footnote{There exist some
subtleties in the identification for $\lambda \neq 1$
\cite{Park:0906}. But here we consider the $\lambda$-deformed
Einstein-Hilbert action $ S_{\lambda \rm EH} = ({c^4}/{16 \pi G_N })
\int dt d^3 x \sqrt{g}N [{c^{-2}}(K_{ij}K^{ij}-\lambda
K^2)+R^{(3)}-{2 {c^{-2}}\Lambda} ] $ following
\cite{Horava,Park:0906,Park:0910a,Park:0910b}.
%, in the convention of \cite{Ryde}.
}. More generally,
%We should point out that at IR limit the action (\ref{Action}) can
%be reduced to Einstein-Hilbert action with the cosmological
%constant($\Lambda$) for $\varphi=\varphi_0=const$ and the following
%conditions:
%\begin{eqnarray}
%\frac{2}{\kappa^2}=\frac{1}{16\pi G_N c},~~~\varphi_0=\frac{c}{16\pi
%G_N},~~~V(\varphi_0)=\frac{2\Lambda c}{16\pi G_N},\label{Elimit}
%\end{eqnarray}
%where $G_N$ is the gravitational constant and $c$ is the speed of
%light. We should point out that from (\ref{Elimit}) Einstein limit
%with $\lambda=1$ can be achived independent of the cosmological
%constant term($\Lambda$).
we can also check that in the case of the parameters given by
\begin{eqnarray}
&&\varphi_0=\left[\frac{\kappa^2\mu^2(\Lambda_W-\omega)}{8(1-3\lambda)}\right]^{1/8},
~~~
%V_{\nu}(\varphi_0)=\frac{3\kappa^2\mu^2\Lambda_{W}^2}{8(1-3\lambda)}=\nu
%\varphi_0^{12},~~
\nu=\left[ \frac{72 (1-3 \lambda) \Lambda_W^4}{\kappa^2 \mu^2
(\Lambda_W-\omega)^3} \right]^{1/2},\nn\\
&&\beta_1=\frac{|\kappa\mu|(1-4\lambda)}{16}
\left[\frac{2}{(\Lambda_W-\omega)(1-3\lambda)}\right]^{1/2},
~~\beta_2=-\frac{|\kappa\mu|}{2}\left[\frac{1-3\lambda}{2(\Lambda_W-\omega)}\right]^{1/2}\nn\\
&&\beta_3=-\frac{\kappa^2}{4W^4},~~\beta_4=\frac{5\kappa^2}{4W^4},
~~\beta_5=-\frac{3\kappa^2}{2W^4},
~~\beta_6=-\frac{\kappa^2}{2W^4},~~\beta_7=\frac{3\kappa^2}{16W^4},
\end{eqnarray}
 the ``background'' action for the scalar field $S(\varphi_0)$ can
be reduced to the IR-modified HL action (without parity violation),
\begin{eqnarray}
&&S_{HL}=\int dtd^3x\,
\sqrt{g}N\left\{\frac{2}{\kappa^2}(K_{ij}K^{ij} -\lambda
K^2)+\frac{\kappa^2\mu^2(\Lambda_W-\omega)
}{8(1-3\lambda)}R-\frac{3\kappa^2\mu^2\Lambda_W^2}{8(1-3\lambda)}\right.\nn\\
&&\hspace*{7em}\left.
-\frac{\kappa^2}{2W^4}C_{ij}C^{ij}+\frac{\kappa^2\mu^2
(1-4\lambda)}{32(1-3\lambda)}R^2
-\frac{\kappa^2\mu^2}{8}R_{ij}R^{ij}\right\},\label{action}
\end{eqnarray}
 where $\omega$ is an arbitrary constant parameter which breaks
the detailed balance softly in IR \cite{Horava,Keha,Park:0905}. But,
we stress that there is no symmetry which can gauge away the scalar
fluctuation mode generally, i.e., for arbitrary parameters $\xi_1,
\xi_2, \xi_3$
 the pathological scalar graviton problem may be different as one can see
in the following sections. And this scalar mode carries the Lorentz
violation effect even in IR, in addition to the IR Lorentz violation
parameter $\lambda$ in HL gravity; the action (\ref{Action}) can be
considered as a (power-counting) renormalizable, Lorentz-violating
scalar-tensor theory. Note also that in the action (\ref{Action}) we
have $3$ new parameters, $\xi_{1\sim3}$ compared to HL gravity, and
the coefficient of the Cotton term $ -\kappa^2 C_{ij}C^{ij}/2W^4$ in
(\ref{action}) is determined from specific values of
$\beta_{3\sim7}$.

\section{Scalar graviton mode %of the metric
in the quadratic action}

~~~~~In order to check explicitly the scalar graviton problem in the
Minkowski background, we first consider the terms that contribute to
the quadratic action for the linear perturbation as
%(after re-scaling $\varphi^8\rightarrow 2\varphi^8/\kappa^2$)
\begin{eqnarray} &&\hspace*{-2em}S=\int \,dt
d^3xN\sqrt{g}~\Bigg\{ \frac{2}{\kappa^2}
\left(\tilde{B}_{ij}\tilde{B}^{ij} -\lambda
\tilde{B}^2\right)-V_{\nu}(\varphi)+\varphi^{8}\tilde{R}+ \beta_1
%\sqrt{\frac{\kappa^2}{2}}
\varphi^{4}\tilde{R}^2
+\beta_2
%\sqrt{\frac{\kappa^2}{2}}
\varphi^4(\tilde{R}_{ij})^2
\nn\\
&&\hspace*{13em}
+\beta_6
%\frac{\kappa^2}{2}
\bar{\nabla}_{i}\tilde{R}_{jk}
\bar{\nabla}^{i}\tilde{R}^{jk}
+\beta_7
%\frac{\kappa^2}{2}
(\bar{\nabla}_i
\tilde{R})^2\Bigg\}.\label{2action}
\end{eqnarray}
Then we %first introduce
 consider the
%general
 following scalar perturbations of the metric
%, up to second order \cite{mukhanov}
and the matter field for the Minkowski background with $\Lambda=0$
[$\varphi_0=({c^4}/{16\pi G_N})^{1/8},~V_{\nu}(\varphi_0)={2\Lambda
c^2}/{16\pi G_N}=0$], up to the linear order ($\triangle\equiv
\partial_i \partial^i$)
%$\varphi$ ($V(\varphi_0)=0$)
\begin{eqnarray}
&&N=1+\phi
%-\frac{1}{2}\phi^2+\frac{\partial_iB\partial_iB}{2}
,~~N_i =
 \partial_iB, %\label{pmetric1}\\
~~g_{ij}=(1-2\psi)\delta_{ij}+2
\frac{\partial_i\partial_j}{\triangle}E, %\nn \label{pmetric2}\\
%&&g^{ij}=\delta_{ij}+2\psi\delta_{ij}-2\partial_i\partial_jE
%+4\psi^2\delta_{ij}+4\partial_k\partial_iE\partial_j\partial_kE
%-8\psi\partial_i \partial_j E ,\nn\label{pmetric3}
%&&\sqrt{g}=1-3\psi
%+\frac{3}{2}\psi^2
%+\partial_i\partial_iE
%+\frac{1}{2} \partial_i\partial_iE\partial_j\partial_jE
%-\partial_j\partial_iE\partial_i\partial_jE-\partial_i\partial_iE\psi
%\nn %\label{pmetric4},\\
~~\varphi=\varphi_0 +\tilde{\varphi}.\label{pphi}
\end{eqnarray} %(Mukhanov).
%Note that the length element has the same form as in the first order
%perturbation.
Substitution of the above perturbations into the action
(\ref{2action}) leads to the following quadratic action (by adopting
the convention $\kappa^2 = 2$)
\begin{eqnarray}
&&\hspace*{-1.2em}S^{(2)}=\int \,dt d^3x\left\{
-6\dot{\psi}^2+16\xi_1
\dot{\psi}\frac{\dot{\tilde{\varphi}}}{\varphi_0}-24\xi_1^{2}\frac{\dot{\tilde{\varphi}}^2}{\varphi_0^{2}}+4\psi\triangle
\dot{B}-4\psi
\ddot{E}-8\xi_1\triangle\dot{B}\frac{\tilde{\varphi}}{\varphi_0}
-8\xi_1\ddot{\psi}\frac{\tilde{\varphi}}{\varphi_0}\right.\nn\\
&&
%\left.
\hspace*{6em}
+8\xi_1\ddot{E}\frac{\tilde{\varphi}}{\varphi_0}+(1-\lambda)\left(3\dot{\psi}-6\xi_1
\frac{\dot{\tilde{\varphi}}}{\varphi_0}+\triangle B-\dot{E}\right)^2
%\right.
\nn\\
&&%\left.
\hspace*{6em}
-2\varphi_0^{8}\left(\psi\triangle\psi-2\phi\triangle\psi-16\frac{\tilde{\varphi}}{\varphi_0}\triangle\psi\right)-8\xi_3
\varphi_0^{8}\left(\phi+7\frac{\tilde{\varphi}}{\varphi_0}\right)\frac{\triangle\tilde{\varphi}}{\varphi_0}
%\right.
\nn\\
&& %\left.
+\beta_1
%\sqrt{\frac{\kappa^2}{2} }
\varphi_0^{4}\left(4\triangle\psi-8\xi_3\frac{\triangle\tilde{\varphi}}{\varphi_0}\right)^2
%-\zeta
+\beta_2
%\sqrt{\frac{\kappa^2}{2} }
\varphi_0^4\left(\partial_i\partial_j\psi+\delta_{ij}\triangle\psi
-2\xi_3\frac{\partial_i\partial_j\tilde\varphi}{\varphi_0}-2\xi_3\delta_{ij}\frac{\triangle\tilde\varphi}{\varphi_0}\right)^2
%\right.
\nn\\
&&%\hspace*{-1em}
\left.-6\beta_6
%\frac{\kappa^2}{2}
\left(\psi-2\xi_3\frac{\tilde{\varphi}}{\varphi_0}\right)
\triangle^3\left(\psi-2\xi_3\frac{\tilde{\varphi}}{\varphi_0}\right)
-16\beta_7
%\frac{\kappa^2}{2}
\left(\psi-2\xi_3\frac{\tilde{\varphi}}{\varphi_0}\right)
\triangle^3\left(\psi-2\xi_3\frac{\tilde{\varphi}}{\varphi_0}\right)
 \right\}. \label{qaction}
\end{eqnarray}

Varying the quadratic action with respect to $\phi$ and $B$, we
obtain the (local) Hamiltonian and momentum constraints (assuming
regular boundary conditions)\footnote{This corresponds to varying
the action with respect to $\triangle \phi$ and $\triangle B$,
instead of $\phi$ and $B$, with the appropriate integration by
parts.}
%The equations of motion for $\phi$ and $B$ are obtained by
\begin{eqnarray}
&& %\triangle
\psi-2\xi_3\frac{
%\triangle
\tilde{\varphi}}{\varphi_0}=0, \label{phieq}\\
&&
%3\dot{\psi}-6\xi_1 \frac{\dot{\tilde{\varphi}}}{\varphi_0}
\frac{1-3 \lambda}{1-\lambda}\left(\dot{\psi}
-2\xi_1\frac{\dot{\tilde{\varphi}}}{\varphi_0}\right)+\triangle
B-\dot{E}=0. \label{Beq}
\end{eqnarray}
 Note that the above action does not
%include
have the contributions
%to
for the higher
%order
derivative terms
%in Eq.(\ref{qaction})
due to the Hamiltonian constraints (\ref{phieq})\footnote{For the
usual tensor graviton modes, however, we have the same
higher-derivative contributions as in HL gravity such that the
(power-counting) renormalizability is not lost with our new
extension. And even for the scalar graviton mode, it is generally
expected that the renormalizability can be maintained again due to
non-linear corrections: The cancelation of higher-derivative terms
is peculiar to the linear perturbation but not generally true in
higher-order perturbations.}, in contrast to the HL case, and nor
the $\xi_2$ dependence. Note also that, for non-vanishing $\xi_3$,
the momentum constraint (\ref{Beq}) further reduces to
\begin{eqnarray}
\frac{1-3 \lambda}{1-\lambda}\left( 1-\frac{\xi_1}{\xi_3}\right)
\dot{\psi}+\triangle B-\dot{E}=0. \label{Beq2}
\end{eqnarray}

By substituting $\tilde{\varphi}$ and $\triangle B-\dot{E}$ with
$\psi$, from the constraints (\ref{phieq}) and (\ref{Beq2}) with the
appropriate integrations by parts, the above quadratic action
becomes
\begin{eqnarray}
S^{(2)}=2
%\frac{4}{\kappa^2}
\int \,dt d^3x\left\{-\frac{1}{c^2_{\psi}} \dot{\psi}^2+
\frac{1-\xi_3}{\xi_3} c^2 \psi\triangle\psi\right\},\label{quad2}
%\int \,dt d^3x\left\{A_1 \dot{\psi}^2+A_2
%\psi\triangle\psi\right\},\label{quad2}
\end{eqnarray}
where
\begin{eqnarray}
c^2_{\psi}=\frac{1-\lambda}{3\lambda-1}\left(\frac{\xi_1}{\xi_3}-1\right)^{-2}.
\label{c_psi}
%\left(\frac{\xi_3}{\xi_1-\xi_3}\right)^2.
%&&A_1=\frac{2(3\lambda-1)}{\lambda-1}\left(\frac{\xi_1}{\xi_3}-1\right)^2,
%~~A_2=2\frac{1-\xi_3}{\xi_3}
\end{eqnarray}
 We note that, when $\xi_3 \rightarrow \infty $, we obtain the
same result as in the HL gravity %for the projectable case
\begin{eqnarray}
S^{(2)}_{HL}=2 %\frac{4}{\kappa^2}
\int \,dt d^3x\left\{-\frac{1}{c^2_{HL}} \dot{\psi}^2- c^2
\psi\triangle\psi\right\},\label{quad_HL}
%\int \,dt d^3x\left\{A_1 \dot{\psi}^2+A_2
%\psi\triangle\psi\right\},\label{quad2}
\end{eqnarray}
with $c^2_{HL}=({1-\lambda})/({3\lambda-1})$ if we ignore the
higher-derivatives terms which were kept there.\footnote{This would
be clear in the action (\ref{2action}), where the matter
perturbation $\tilde{\varphi}$ is decoupled from the gravity part in
the $\xi_3 \rightarrow \infty$ limit; if we consider first the
$\xi_3 \rightarrow \infty$ limit before implementing the Hamiltonian
constraint (\ref{phieq}), we can recover the higher-derivative terms
also.} In this case, it is known that scalar graviton $\psi$ has
several pathological behaviors: $\psi$ would be either unstable when
$c^2_{HL} <0$ or be a ghost when $c^2_{HL}>0$; moreover, there are
strongly coupled interactions for $c_{HL}\rightarrow 0$ $(\lambda
\rightarrow 1)$, i.e., the perturbation around the Minkowski
background can not be defined in the Einstein gravity limit
$(\lambda \rightarrow 1)$ \cite{koyama}.

But in our case, we can cure the instability/ghost problem with
\footnote{The ghost can be also avoided with $\lambda <1/3$ but we
do not consider this possibility here since the Einstein gravity
with $\lambda=1$ can not be obtained.}
%We should point out two comments by comparing BPS gravity
%\cite{sotiriou2,Blas}. First when $\xi_1=0$ or $\xi_1=\eta$, and
%$\xi_3=\eta/2$ the above result is in agreement with
%\cite{sotiriou2,Blas}. Here $\eta$ is the coefficient of
%$\nabla_iN\nabla^iN/N^2$ in BPS gravity. As if in BPS gravtiy the
%presence of $\nabla_iN\nabla^iN/N^2$ term contributes for the scalar
%mode to be neither a ghost nor classically unstable, in our case
%$\nabla_i\nabla^i\varphi/\varphi$ term plays such a role in the
%region:
\begin{eqnarray}
c^2_{\psi}<0 ~~(1<\lambda ),~~0<\xi_3<1 . \label{condition1}
\end{eqnarray}
This situation may be compared with the BPS extension
%gravity
\cite{sotiriou2,Blas}, where the non-projectable lapse function
$N(t, {\bf x})$ becomes a dynamical scalar field by adding the
potential term $V=\eta
{\nabla_iN\nabla^iN}/{N^2}$+(higher-derivative terms). In this case
the resulting scalar graviton action in the quadratic order becomes
\begin{eqnarray}
S^{(2)}_{BPS}=2 %\frac{4}{\kappa^2}
\int \,dt d^3x\left\{-\frac{1}{c^2_{HL}} \dot{\psi}^2-
\frac{\eta-2}{\eta} c^2 \psi\triangle\psi\right\},\label{quad_BPS}
\end{eqnarray}
and the above problems of the instability/ghost can be also cured
for $c^2_{HL}<0,~0<\eta <2.$ But regarding the strong coupling
problem which persists in the BPS extension
%gravity
still
\cite{sotiriou2,Kimp,Soti:1103,blas3,blas4}, we can cure this
problem also in our construction, as will be shown in the next
section.

We finally remark that for the special case of $\xi_3=\xi_1$, the
$\dot{\psi}^2$ term is disappearing even though the spatial
derivatives term remains. This means that there is no dynamical
scalar graviton at the quadratic order.
%Secondly, note that in our case we have one parameter additionally.
%Most of all in the case of
%\begin{eqnarray}
%(\xi_1/\xi_3-1)\sim (\lambda-1)^m~ (m\geq1/2),\label{condition}
%\end{eqnarray}
%$A_1$ behaves regularly without any singular point at
%$\lambda\rightarrow 1$ limit. In the next section, we will show that
%the additional parameter plays also the role of curing the strong
%coupling problem. Note also that there is no contribution from the
%higher order derivative terms.

\section{Strong coupling in the cubic action}

~~~~~We now turn to the cubic-order perturbation of the action to
check whether the strong coupling problem can be resolved in our
framework. Since the issue is about the non-linear perturbation at
low energies (IR), it is enough to
%To this end let us first
consider the action at IR limit:
% without the potential term:
\begin{eqnarray}
S^{(IR)}=%\frac{2}{\kappa^2}
\int \,dt d^3xN\sqrt{g}~\left\{\tilde{B}_{ij}\tilde{B}^{ij} -\lambda
\tilde{B}^2+\varphi^{8} \tilde{R}
%\left(R-8\xi_3 \frac{\nabla_{i}\nabla^{i}\varphi}{\varphi}\right)
-V_{\nu}(\varphi)\right\}.
%\nn
\end{eqnarray}
In order to study
%compare with
the cubic order interaction terms
% in the literature
% it is more practical to
 we consider the non-linear scalar perturbations
around the Minkowski background ($V_{\nu}(\varphi_0)=0$), without
loss of generality\footnote{
%By setting the expansions
%$\phi=\phi^{(1)}+\phi^{(2)}+ \cdots,~
%\psi=\psi^{(1)}+\psi^{(2)}+\cdots,~B=B^{(1)}+B^{(2)}+ \cdots,~
% \tilde{\varphi}=\tilde{\varphi}^{(1)}+\tilde{\varphi}^{(2)}+ \cdots$
One can transform these exponential-type perturbations to the
 other more general power-type perturbations like as $\bar{N}=1+\bar{\phi}+c_2
 \bar{\phi}^2+\cdots,~\bar{g}_{ij}=1-2 \bar{\psi}+d_2 \bar{\psi}^2 + \cdots$
 by setting $\phi=\phi^{(1)}+\phi^{(2)}+ \cdots,~
\psi=\psi^{(1)}+\psi^{(2)}+\cdots$
%,~B=B^{(1)}+B^{(2)}+ \cdots,
%~\tilde{\varphi}=\tilde{\varphi}^{(1)}+\tilde{\varphi}^{(2)}+ \cdots$
and $\bar{\phi}=\bar{\phi}^{(1)}+\bar{\phi}^{(2)}+ \cdots,~
\bar{\psi}=\bar{\psi}^{(1)}+\bar{\psi}^{(2)}+\cdots$.}
%of the metric
as follows \cite{sotiriou2,koyama,Wang}
\begin{eqnarray}
N=e^{\phi},~~N_i=\partial_iB,~~g_{ij}=e^{-2\psi}\delta_{ij}\label{pertur},
~~\varphi=\varphi_0+\tilde{\varphi}.
\end{eqnarray}
 Here we choose the $E=0$ gauge in the most general scalar
perturbation (\ref{pphi}) to simplify the computations.
%(see Appendix A for the gauge equivalence with the scalar
%perturbation of Eq.(\ref{pphi}))
After some manipulations
%through the perturbations (\ref{pertur})
one can find the
%$3$rd
 cubic-order action given by
\begin{eqnarray}
&&\hspace*{-1.2em}S^{(3)}=\nn
\\ &&
%\frac{2}{\kappa^2}
\int \,dt
d^3x\Bigg\{-16\varphi_0^7{\tilde\varphi}(\partial\psi)^2+2\varphi_0^8\psi(\partial\psi)^2
-2\phi\varphi_0^8(\partial\psi)^2+112\varphi_0^6{\tilde\varphi}^2\partial^2\psi
-32\varphi_0^7\psi{\tilde\varphi}\partial^2\psi\nn\\
&&+32\varphi_0^7\phi{\tilde\varphi}\partial^2\psi
+2\varphi_0^8\phi^2\partial^2\psi+2\varphi_0^8\psi^2\partial^2\psi-4\varphi_0^8\phi\psi\partial^2\psi
-8\xi_3\Big(\frac{\varphi_0^7}{2}\phi^2\triangle{\tilde\varphi}+\frac{\varphi_0^7}{2}\psi^2\triangle{\tilde\varphi}\nn\\
&&-\phi\varphi_0^7\psi\triangle{\tilde\varphi}-7\varphi_0^6{\tilde\varphi}\psi\triangle{\tilde\varphi}
+7\varphi_0^6\phi\tilde\varphi\triangle{\tilde\varphi}+21\varphi_0^5{\tilde\varphi}^2\triangle{\tilde\varphi}
+\varphi^7\psi\partial_i
\tilde\varphi\partial_i\psi-\varphi_0^7\phi\partial_i\psi\partial_i\tilde\varphi
\nn\\
&&-7\varphi_0^6\tilde\varphi\partial_i\tilde\varphi\partial_i\psi\Big)
-9(1-3\lambda)\psi\dot{\psi}^2-2(1-3\lambda)\psi\dot\psi\triangle
B-2(1-3\lambda)\dot\psi\partial_k\psi\partial_k
B\nn\\
&&-2(1-\lambda)\triangle B\partial_k\psi\partial_k B
+\psi\partial_i\partial_j B\partial_i\partial_j
B+4\partial_i\partial_j B\partial_i
B\partial_j\psi-\lambda\psi(\triangle
B)^2-3(1-3\lambda)\phi\dot\psi^2\nn\\
&&-2(1-3\lambda)\phi\dot\psi\triangle B-\phi\partial_i\partial_j
B\partial_i\partial_j B+\lambda\phi(\triangle
B)^2-4(1-3\lambda)\varphi_0^{-1}\Big(-6\xi_1\psi\dot\psi\dot{\tilde\varphi}\nn\\
&&-3\xi_2\dot\psi\partial_i\tilde\varphi\partial_i B
-\xi_1\partial_i B\partial_i \psi\dot{\tilde\varphi}-\xi_2\triangle
B\partial_i\tilde\varphi\partial_i
B-3\xi_1\phi\dot\psi\dot{\tilde\varphi}-\xi_1\phi\triangle
B\dot{\tilde\varphi}-3\xi_1\psi\dot\psi\dot{\tilde\varphi}\nn\\
&&-\xi_1\psi\triangle
B\dot{\tilde\varphi}-3\xi_1\varphi_0^{-1}\tilde\varphi\dot\psi\dot{\tilde\varphi}
-\xi_1\varphi_0^{-1}\triangle B\dot{\tilde\varphi}\tilde\varphi
\Big)
+12(1-3\lambda)\varphi_0^{-2}\left(-2\xi_1\xi_2\dot{\tilde\varphi}\partial_i\tilde\varphi\partial_i
B\right.\nn\\
&&\left.-\xi_1^2\phi\dot{\tilde\varphi}^2-3\xi_1^2\psi\dot{\tilde\varphi}^2
-2\xi_1^2\varphi_0^{-1}\tilde\varphi\dot{\tilde\varphi}^2\right)\Bigg\}
\label{cubic}
\end{eqnarray}
%By
Using the first-order Hamiltonian and momentum constraints
%Eqs.
(\ref{phieq}), (\ref{Beq}) (or (\ref{Beq2})) obtained in the
previous section\footnote{In order to compute the cubic-order
interaction one only needs to consider the constraints for the
perturbations of $N$ and $N_i$ to the first order \cite{Mald:0210}.
More generally, for the $n$'th-order interactions, one only needs to
consider the $(n-2)$'th order \cite{Seer}.}, the above action
(\ref{cubic}) reduces to
%can be written as
\begin{eqnarray}
%&&\hspace*{-1.3em}
S^{(3)}&=&
%\nn\\&&
%\frac{4}{\kappa^2}
 2\int \,dt d^3x
\left\{\left(-1+\frac{5}{\xi_3}-\frac{7}{\xi_3^2}\right)c^2
%\varphi_0^8
 {\psi}(\partial_i{\psi})^2
-\frac{2}{c^4_{\psi}}\left(\frac{\xi_1}{\xi_3}-1\right)^{-2}
\left(\frac{\xi_2}{\xi_3}-1\right)
\dot{\psi}\partial_i{\psi}\partial_i\left(\frac{\dot{\psi}}{\triangle}\right)
\right.\nn\\
&& \hspace*{6em}
\left.-\frac{3}{2c^4_{\psi}}\left(\frac{\xi_1}{\xi_3}-1\right)^{-2}
\psi\left(\frac{\partial_i\partial_j}{\triangle}\dot{\psi}\right)^2+
\left[\frac{3}{2c^4_{\psi}}\left(\frac{\xi_1}{\xi_3}-1\right)^{-2}
\right.\right.\nn\\
&&\hspace*{14em} \left.\left.
-\frac{1}{c^2_{\psi}}\left(\frac{\xi_1}{\xi_3}-1\right)^{-1}
\left(3-3\frac{\xi_1}{\xi_3}-\frac{\xi_1}{\xi_3^2}\right)\right]
{\psi}\dot{\psi}^2\right\}.\label{cubic1}
\end{eqnarray}
Note again that when %if
$\xi_3\rightarrow\infty$ the above action can be reduced to the
cubic action in the HL gravity
%, for the projectable case,
as
%(for $\varphi_0=1$)
\cite{koyama,sotiriou2}
\begin{eqnarray}
&&\hspace*{-1.3em} S^{(3)}_{HL}= 2
%\frac{4}{\kappa^2}
\int \,dt d^3x \Bigg\{-
%\frac{c^4}{16 \pi G_N}
 c^2{\psi}(\partial_i{\psi})^2
+\frac{2}{c^4_{HL}}\dot{\psi}\partial_i{\psi}\partial_i\left(\frac{\dot{\psi}}{\triangle}\right)\nn\\
&&\hspace*{10em}+\frac{3}{2}\left[-\frac{1}{c^4_{HL}}\psi\left(\frac{\partial_i\partial_j}{\triangle}\dot{\psi}\right)^2
+\frac{2c^2_{HL}+1}{c^4_{HL}}{\psi}\dot{\psi}^2\right]\Bigg\}.
%\nn
\end{eqnarray}

Now, in order to discuss the strong coupling problem we use the
canonically normalized variable $\hat{\psi}= \bar{M}_{Pl}
\psi/|c_{\psi}|$
%($M^2_{Pl} \equiv 4/\kappa^2$)
 (by recovering $2/\kappa^2
%=c^2/(16 \pi G_N)
=c M_{Pl}^2/16 \pi \hbar \equiv \bar{M}_{Pl}^2/2)$ such that the
quadratic action (\ref{quad2}) becomes
\begin{eqnarray}
S^{(2)}= \int \,dt d^3x\left\{\dot{\hat{\psi}}^2+
\frac{1-\xi_3}{\xi_3}~ c^2
|c_{\psi}|^2\hat{\psi}\triangle\hat{\psi}\right\}.\label{quad3}
\end{eqnarray}
 Then, the cubic action (\ref{cubic1}) becomes
\begin{eqnarray}
&&\hspace*{-1.3em}S^{(3)}= \nn\\
&&\hspace*{-1.3em}\frac{1}{ \bar{M}_{Pl}}
%\left(\frac{8 \pi \hbar}{c}\right)^{1/2}
\int \,dt d^3x
\left\{\left(-1+\frac{5}{\xi_3}-\frac{7}{\xi_3^2}\right)
%\varphi_0^8
c^2 |c_{\psi}|^3 \hat{\psi}
 (\partial_i \hat{\psi})^2
-\frac{2}{|c_{\psi}|}\left(\frac{\xi_1}{\xi_3}-1\right)^{-2}\left(\frac{\xi_2}{\xi_3}-1\right)
\dot{\hat{\psi}}\partial_i\hat{\psi}\partial_i
\left(\frac{\dot{\hat{\psi}}}{\triangle}\right) \right.\nn\\
&&\hspace*{7em}\left.-\frac{3}{2|c_{\psi}|}\left(\frac{\xi_1}{\xi_3}-1\right)^{-2}
\hat{\psi}\left(\frac{\partial_i\partial_j}{\triangle}\dot{\hat{\psi}}\right)^2+
\left[\frac{3}{2|c_{\psi}|}\left(\frac{\xi_1}{\xi_3}-1\right)^{-2}\right.\right.\nn\\
&&\hspace*{15em}\left.\left.+|{c_{\psi}}|\left(\frac{\xi_1}{\xi_3}-1\right)^{-1}
\left(3-3\frac{\xi_1}{\xi_3}-\frac{\xi_1}{\xi_3^2}\right)\right]
\hat{\psi}\dot{\hat{\psi}}^2\right\}.\label{cubic2}
\end{eqnarray}
 Note that all the terms but the first term scale as
$c_{\psi}^{-1}$ and so there is strong coupling for $\lambda
\rightarrow 1$ since $c_{\psi}\rightarrow 0$ naively, from
(\ref{c_psi}): All the cubic interaction terms that have the time
derivatives of $\psi$ blow up in that limit. However in our
construction, due to the presence of another coupling $\xi_1$, which
would be running in principle, this strong coupling problem can be
cured by the ``fine tuning '' in the limit of $\lambda \rightarrow
1$. If
\begin{eqnarray}
(\xi_1-\xi_3)\sim (\lambda-1)^s~~ (s \leq -1/2),\label{condition2}
\end{eqnarray}
 then
\begin{eqnarray}
|c_{\psi}|\sim \frac{\sqrt{\lambda-1}}{ |\xi_1-\xi_3|}
\sim(\lambda-1)^{1/2-s},
\end{eqnarray}
% where $c^2=(\lambda-1)/(1-3\lambda)$.
 and the troublesome interactions which scale as
\begin{eqnarray}
\frac{1}{|c_{\psi}|} \left( \frac{\xi_1}{\xi_3}-1 \right)^{-2}
\sim\frac{1}{(\lambda-1)^{s+1/2}}
\end{eqnarray}
can be made to be regular. The cubic action becomes finite as
\begin{eqnarray}
\hspace*{-1.3em}S^{(3)}\sim \frac{1}{\bar{M}_{Pl}} \int \,dt d^3x
\left\{ -{2}\left(\frac{\xi_2}{\xi_3}-1\right)
\dot{\hat{\psi}}\partial_i\hat{\psi}\partial_i
\left(\frac{\dot{\hat{\psi}}}{\triangle}\right) %\right.\nn\\
%\hspace*{-2.6em}\left.
-\frac{3}{2}
\hat{\psi}\left(\frac{\partial_i\partial_j}{\triangle}\dot{\hat{\psi}}\right)^2+
%\left\{
\frac{3}{2}
%\right\}
\hat{\psi}\dot{\hat{\psi}}^2\right\}.\label{cubic3}
\end{eqnarray}
 for $s=-1/2$ or vanishing %$S^{(3)}\sim 0$
for $s<-1/2$. On the other hand, in this case, the quadratic action
(\ref{quad3}) becomes
\begin{eqnarray}
S^{(2)}\sim  \int \,dt d^3x~\dot{\hat{\psi}}^2\label{quad4}
\end{eqnarray}
 such that there is no ghost/instability problem either. Note that
the condition (\ref{condition2}) for the absence of strong coupling
for $\lambda \rightarrow 1$ is consistent with the condition
(\ref{condition1}) for the absence of ghost/instability for $\lambda
>1$; this is in contrast to BPS gravity case \cite{Soti:1103}.
Moreover, for the special case of $\xi_1=\xi_3$ in the action
(\ref{cubic1}), all the cubic terms that have the time-derivatives
of $\psi$ vanish and
% In this action, as was done in Sec.3, we comment on two cases in
%order.\\
%Case I: $\xi_1=\xi_3$\\
%\\
%In this case,
the above cubic action (\ref{cubic2}) reduces to
\begin{eqnarray}
S^{(3)}=
%\frac{4}{\kappa^2}
\frac{1}{ \bar{M}_{Pl}} \int \,dt d^3x
\left\{\left(-1+\frac{5}{\xi_3}-\frac{7}{\xi_3^2}\right) c^2
|c_{\psi}|^3 \hat{\psi} (\partial_i \hat{\psi})^2
%\varphi_0^8\psi(\partial_i\psi)^2
\right\}.
\label{caction1}
\end{eqnarray}
%From the above action (\ref{caction1}), we can find
 such that there is no strong coupling problem either, unless
$\xi_3=0$.
%, especially when $\varphi_0=0$, the cubic terms can be
%completely disappeared.
Note that in this case, the quadratic action (\ref{quad2}) did not
have
%did not include any
 the time-derivative term either.
%Note also that in the
%action (\ref{caction1}), the interaction terms with time derivatives
%are also absent in contrary to the projectable Horava-Lifshitz
%gravity \cite{koyama,blas2}.

%Case II: $\xi_1\neq\xi_3$\\
%\\
%Let us first consider the limit of $\lambda\rightarrow 1$. From the
%above action (\ref{cubic1}) we can read that $(\xi_1/\xi_3-1)$ and
%$(3\lambda-1)/(\lambda-1)$ have the same order in the 2nd, 3rd, 4th
%terms. In order to have no singular points at $\lambda\rightarrow1$
%limit one can easily choose the condition as
%\begin{eqnarray}
%(\xi_1/\xi_3-1)\sim (\lambda-1)^l~~ (l\geq1).\label{condition2}
%\end{eqnarray}
%Note that by comparing with (\ref{condition}) found in Sec.3. one
%can show that the above region (\ref{condition2}) makes our system
%possible to be free from scalar graviton and strong coupling
%problems.\\
 Finally, we remark that the higher-derivatives terms which have been
ignored in the cubic interaction can not change our conclusion. This
is because they generate only the spatial derivatives of $\psi,
\tilde{\varphi}$, and $E$, not $B$. From the constraints
(\ref{phieq}), (\ref{Beq}) (or (\ref{Beq2})), only $B$ is related to
the time derivative of $\psi$ whose interactions reveal strong
coupling as in the BPS %gravity
case \cite{sotiriou2}.

\section{Conclusion and discussion}

~~~~~We have extended the HL gravity with extra conformal invariance
by introducing an extra scalar field. In the case of the critical
exponent $z=3$, which breaks the equal-footing treatment of space
and time in UV, power counting
renormalizability %is preserved
 can be achieved without the ghost problem for the transverse
traceless graviton modes. Relaxing the exact Weyl symmetry, we
considered an action with three new coupling parameters
$\xi_1,\xi_2, \xi_3$ which breaks the {\it local} anisotropic Weyl
symmetry but still preserves residual {\it global} Weyl invariance.
With a constant scalar field and %the
$\xi_3 \rightarrow \infty$
limit,
%particular %choices
%{\bf values} of the {\bf new coupling} parameters,
it reduces to the
%Horava-Lifshitz
 HL gravity, but it generally have some more degrees of
freedom to cure the pathologies of the scalar graviton. Actually, we
have found that, in the perturbation around the Minkowski
background,
%analysis shows that
both the instability/ghost problem of scalar graviton at the
quadratic order and the strong coupling problem at the cubic order
can be cured by the appropriate fine tuning of the (running)
couplings as $\lambda \rightarrow 1_{+},~0< \xi_3
<1,~\xi_1-\xi_3\sim (\lambda-1)^s~~ (s \leq -1/2)$.
%does not emerge up to cubic order as well as
%quadratic order.
This implies that the scalar matter field $\tilde{\varphi}$, which
drives the scalar graviton $\psi$ as in (\ref{phieq}), (\ref{Beq})
(or (\ref{Beq2})), regularizes the strongly-coupled cubic
interactions and makes the scalar graviton healthy in the
%cubic
quadratic propagation as well. This is in contrast to BPS %gravity
approach
where only one new coupling $\eta$ was relevant in IR and so the
strong coupling problem and the instability/ghost problem can not be
cured simultaneously, unless a new low energy scale below the Planck
scale is introduced.

%It seems that the extension favors the non-projectable case.
%We recall that for $\omega=\omega(t)$ the action preserves local symmetry only for
%$\xi_1=1$ regardless of $\xi_2,\xi_3$ other than the value of 1. In this case,
%{\bf In this work, we have studied the non-projectable case only.
%However, if we consider the projectable case, which is
%projectability can be
%allowed in the theory also, and it turns out that the theory faces
%the pathology problem of ghost or classical instability as in the
%previous attempts.
 For the projectable case, we can not use the Hamiltonian
constraint (\ref{phieq}) anymore and the pure scalar graviton terms
are the same as in the HL gravity. In other words, one has the same
pathologies of scalar graviton as in the (projectable) HL gravity.

 Our new action reduces to Lorentz-violating scalar-tensor
gravity theory at low energies. It is known that there is very
strong constraints for the viable scalar-tensor theories
\cite{Tsuj:0803}. It is left as an open problem whether this theory
can be consistent with other observational and local gravity tests
also. It is also challenging to check the closure of algebras from
the full Hamiltonian analysis\footnote{After the completion of this
work, we became aware of the article \cite{Kluson} in which the full
Hamiltonian analysis in the {\it exact} Weyl-invariant, BPS
extension of HL gravity was performed.}.

%\section{Appendix A: Gauge}
%\section{Appendix B: Mukhanov}

\newpage
\section*{Acknowledgments}

~~~~~TM was supported by the National Research Foundation of
Korea(NRF) grant funded by the Korea government(MEST) through the
Center for Quantum Spacetime(CQUeST) of Sogang University with grant
number 2005-0049409. PO was supported by the National Research
Foundation of Korea(NRF) grant funded by the Korea government(MEST)
through the Center for Quantum Spacetime(CQUeST) of Sogang
University with grant number 2005-0049409 and by the BSRP through
the National Research Foundation of Korea funded by the MEST
(2011-0026655). MIP was supported by the Korea Research Foundation
Grant funded by Korea Government (MOEHRD) (KRF-2010-359-C00009).

\end{document}